\documentclass[12pt]{emulateapj}

\usepackage{natbib}
\usepackage{apjfonts} 

\newcommand{\etal}{et~al.}
\newcommand{\eg}{e.g.,}

\newcommand{\ie}{i.e.,}

\newcommand{\mum}{$\mu$m}
\newcommand{\muj}{$\mu$Jy}

\newcommand{\hmpc}{\hbox{$h^{-1}$ Mpc}}

\newcommand{\msun}{\hbox{$M_{\odot}$}}

\newcommand{\meanz}{\left<z\right>}
\newcommand{\mtwo}{$M_{\mbox{\scriptsize 200}}$}
\def\spose#1{\hbox to 0pt{#1\hss}}
\def\simlt{\mathrel{\spose{\lower 3pt\hbox{$\mathchar"218$}}
     \raise 2.0pt\hbox{$\mathchar"13C$}}}
\def\simgt{\mathrel{\spose{\lower 3pt\hbox{$\mathchar"218$}}
     \raise 2.0pt\hbox{$\mathchar"13E$}}}

\slugcomment{ApJ Letters, in press}

\shorttitle{Galaxy Cluster Correlation Function to $z \sim 1.5$}
\shortauthors{Brodwin et al.}

\begin{document}


\title{Galaxy Cluster Correlation Function to $z \sim 1.5$ in the IRAC
  Shallow Cluster Survey}


\author{M.~Brodwin\altaffilmark{1,2}, 
A.~H.~Gonzalez\altaffilmark{3},
L.~A.~Moustakas\altaffilmark{1},
P.~R.~Eisenhardt\altaffilmark{1}, 
S.~A.~Stanford\altaffilmark{4,5},
D.~Stern\altaffilmark{1},
and M.~J.~I.~Brown\altaffilmark{6}
}


\altaffiltext{1}{JPL/Caltech, 4800 Oak Grove Dr., Pasadena, CA 91109}
\altaffiltext{2}{NOAO, 950 N.~Cherry Ave., Tucson, AZ 85719}
\altaffiltext{3}{Dept.~of Astronomy, Univ.\ of Florida, Gainesville, FL 32611}
\altaffiltext{4}{University of California, Davis, CA 95616}
\altaffiltext{5}{IGPP, LLNL, Livermore, CA 94550}
\altaffiltext{6}{School of Physics, Monash Univ., Clayton 3800, Victoria, Australia}


\begin{abstract}

  We present the galaxy cluster autocorrelation function of 277 galaxy
  cluster candidates with $0.25 \le z \le 1.5$ in a 7 deg$^2$ area of
  the IRAC Shallow Cluster Survey.  We find strong clustering
  throughout our galaxy cluster sample, as expected for these massive
  structures.  Specifically, at $\meanz = 0.5$ we find a correlation
  length of $r_0 = 17.40^{+3.98}_{-3.10}$ \hmpc, in excellent
  agreement with the Las Campanas Distant Cluster Survey, the only
  other non--local measurement.  At higher redshift, $\meanz = 1$, we
  find that strong clustering persists, with a correlation length of
  $r_0=19.14^{+5.65}_{-4.56}$ \hmpc.  A comparison with high
  resolution cosmological simulations indicates these are clusters
  with halo masses of $\sim 10^{14} M_\odot$, a result supported by
  estimates of dynamical mass for a subset of the sample.  In a stable
  clustering picture, these clusters will evolve into massive
  ($10^{15} M_\odot$) clusters by the present day.
  \end{abstract}





\section{Introduction}

The clustering amplitude of massive galaxy clusters, and in particular
its dependence on richness or cluster mass, is a strong function of
the underlying cosmology.  Recent theoretical work
\citep[\eg][]{majumdar&mohr03,younger06} has demonstrated how cluster
surveys can be ``self--calibrated,'' providing precise simultaneous
constraints on both the cosmology and cluster evolution models.  Given
redshift and approximate mass information for each cluster, reliable
cosmological parameter estimation is feasible even in the presence of
significant, and potentially unknown, evolution in cluster physical
parameters \citep[\eg][]{gladders07}.  The ultimate goal of these
cosmological studies, a measurement of the equation of state of dark
energy with an accuracy competitive with SNe Ia methods, also requires
knowledge of the cluster power spectrum or autocorrelation function
\citep{majumdar&mohr04,wang04}. Such extensive sample characterization
naturally emerges from mid--infrared selected photometric redshift
cluster surveys like the IRAC Shallow Cluster Survey
\citep[ISCS;][hereafter E07]{eisenhardt07}.

Previous studies \citep[\eg][and references therein]{bahcall03} have
produced largely local ($z\la 0.3$) galaxy cluster clustering
measurements with limited baselines for evolutionary studies. The ISCS
provides the opportunity to address these issues in one of the largest
statistical samples of high redshift clusters to date. In this Letter
we present the first measurement of the galaxy cluster autocorrelation
function extending over more than half the cosmic age of the Universe.
Such measurements also allow evolutionary connections between galaxy
clusters and high redshift, highly clustered galaxy populations to be
explored. 

We use a concordance cosmology throughout, with $\Omega_M = 0.3$,
$\Omega_\Lambda = 0.7$, and $H_0 = 70$ km s$^{-1}\!$ Mpc$^{-1}$. For
consistency with previous studies we report distances, including
correlation lengths, in units of comoving \hmpc, with $H_0 = 100 \,h$
km s$^{-1}$ Mpc$^{-1}$.  

\section{IRAC Shallow Cluster Survey}
\label{Sec: sample}

The ISCS is a sample of 335 galaxy clusters spanning $0.1<z<2$ in the
IRAC Shallow Survey \citep[ISS,][]{eisenhardt04} area of the NOAO
Deep-Wide Field Survey (NDWFS, \citealt{ndwfs99}; Jannuzi \etal\ in
prep) in Bo\"otes.  The clusters are selected using a wavelet
detection algorithm which identifies peaks in cluster probability
density maps constructed from accurate photometric redshift
probability functions for 175,431 galaxies brighter than 13.3\muj\ at
4.5\mum\ in a 7.25 deg$^2$ region (\citealt{brodwin06_ISS}, hereafter
B06; E07).  The galaxy photometric redshifts, which are key to the
cluster--finding algorithm, are derived from the joint ISS, NDWFS
(DR3) and FLAMEX \citep{elston06} data sets.

The AGES survey (Kochanek \etal\ in prep) in Bo\"otes provides
spectroscopic confirmation for dozens of clusters at $z\le 0.5$ (E07).
At higher redshift, a multi--year Keck spectroscopic campaign has to
date confirmed 10 $z>1$ clusters (\citealt{stanford05};
\citealt{elston06}; B06; E07).

Extensive masking, described in B06, is implemented to reject areas
suffering from remnant cosmetic artifacts or data quality issues which
could affect the robustness of the clustering measurements.
The final unmasked area is 7.00 deg$^2$ and contains 320 of the 335
galaxy clusters in the full ISCS, of which 277 are within the redshift
range ($0.25 \le z \le 1.5$) considered in this Letter.

\section{Galaxy Cluster Autocorrelation Measurements}
\label{section method}
\subsection{Angular Correlation Function in Redshift Slices}

\begin{figure}[bthp]
\epsscale{1.25}
\hspace*{-0.65cm}
\plotone{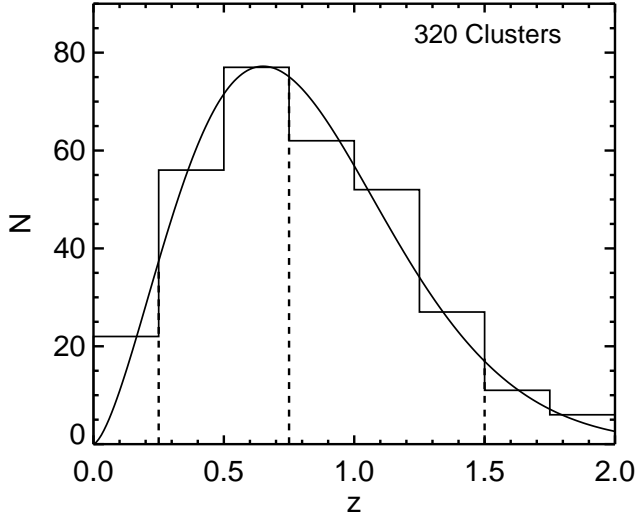}
\caption{Observed redshift distribution of galaxy clusters in the
  ISCS.  The dashed lines illustrate the two redshift bins considered
  here, $0.25 \le z \le 0.75$ and $0.75 < z \le 1.5$, containing 277
  clusters between them.  The curve is a fit to the distribution and
  is used in the Limber deprojection.  This distribution should not be
  used for quantifying number density evolution without careful
  inclusion of selection biases.}
\label{Fig: N(z)}
\end{figure}

The redshift distribution of galaxy clusters is presented in Figure
\ref{Fig: N(z)}.  A fit of the form $N(z) \propto (z/z_0)^{\alpha}
\exp{(-(z/z_0)^{\beta})}$, with $\alpha = 1.5$, $\beta = 1.7$ and $z_0
= 0.7$, is overplotted. Cosmological studies with this observed
redshift distribution must await a detailed characterization of the
survey selection function, to be presented in a forthcoming
paper. Galaxy clusters are split into two similarly populated redshift
bins, $0.25 \le z \le 0.75$ (136 clusters) and $0.75 < z \le 1.5$ (141
clusters), using the best available redshift information (\ie\
spectroscopic redshifts where available, photometric redshifts
otherwise). These bins span the range where the accuracy, $\sigma_z =
0.06(1+z)$, and reliability of the galaxy photometric redshifts have
been demonstrated (B06; see also \citealt{brown06}).

The angular correlation function (ACF) is parametrized here as a
simple power law,
\begin{equation}
\omega(\theta) = A_{\omega}\theta^{-\delta}.
\label{Eq: omega powerlaw}
\end{equation}
This can be deprojected \citep{limber} to yield a measurement of the
real--space correlation length, $r_0(z)$, over the redshift range
spanned by the 2--D sample,
\begin{equation}
r_0^\gamma(z_{\mbox{\scriptsize eff}}) = A_{\omega} \left[\frac{H_0 H_{\gamma}}{c}\, \frac{\int_{z_1}^{z_2} N^2(z)\,[x(z)]^{1-\gamma}\, E(z)\, dz}{[\int_{z_1}^{z_2} N(z)\, dz]^2}\right]^{-1},
\label{Eq: r_0}
\end{equation}
where
\begin{equation}
z_{\mbox{\scriptsize eff}} = \frac{\int_{z_1}^{z_2} z\, N^2(z)\,[x(z)]^{1-\gamma}\, E(z)\, dz}{\int_{z_1}^{z_2} N^2(z)\,[x(z)]^{1-\gamma}\, E(z)\, dz},
\label{Eq: z_eff}
\end{equation}
$\gamma \equiv 1+ \delta$, $H_{\gamma} =
\Gamma(1/2)\,\,\Gamma[(\gamma-1)/2]/\Gamma(\gamma/2)$, $N(z)$ is the
redshift distribution, and $E(z)$ and $x(z)$ describe the evolution of
the Hubble parameter and the comoving radial distance, respectively
\citep[\eg][]{hogg_distance}.  Since the cluster redshift
uncertainties are much smaller than the width of our redshift bins
(E07), they have little impact on the redshift distribution and the
modeling of the spatial correlation function.

\subsection{Results}

\begin{figure}[bthp]
  \epsscale{1.2}
  \plotone{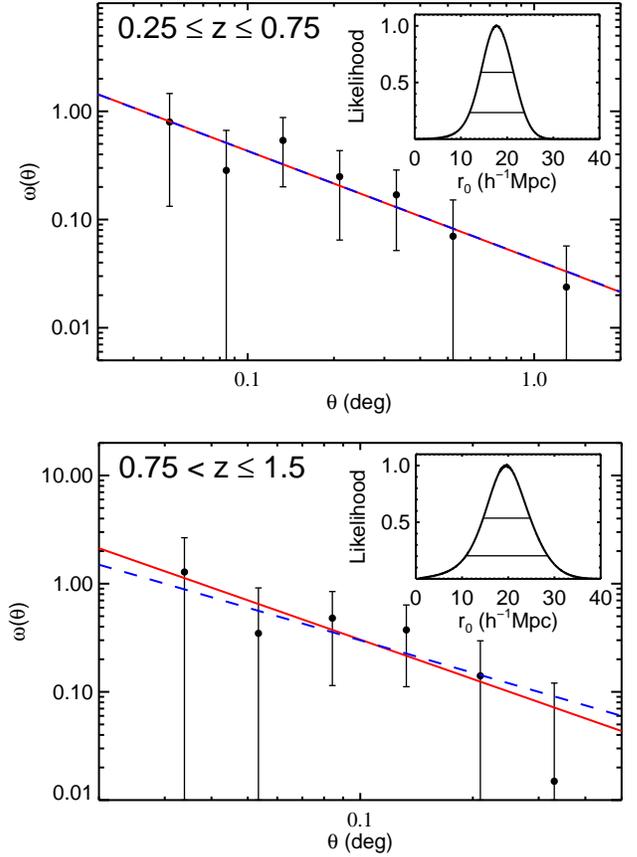}  
  \caption{Angular correlation functions in redshift bins at $0.25 \le z
    \le 0.75$ ({\it top}) and $0.75 < z \le 1.5$ ({\it bottom}).  The
    error bars are estimated via bootstrap resampling.  The red solid
    lines show the best--fit ACFs, with corresponding slopes listed in
    Table \ref{Tab: Fits}.  The blue dashed lines show the fits with a
    fixed slope ($\delta = 1.0$).  Marginalizing over the slope produces
    the likelihood functions in $r_0$ presented in the insets.  The 68\%
    and 90\% confidence intervals are shown.}
  \label{Fig: ACF}
\end{figure}

We calculate the ACF using both the \citet{landy&szalay} and
\citet{hamilton} estimators using 500,000 randoms to ensure a robust
Monte Carlo integration.  The results are nearly identical with both
estimators and we report the results obtained with the latter.  Two
independent fitting techniques are applied to the data. The standard
frequentist (or classical) approach is used to simultaneously fit the
slope, and, through the use of the relativistic Limber equation, the
correlation length, $r_0$. In calculating these correlation lengths we
have adopted the $N(z)$ parametrization shown in Figure \ref{Fig:
  N(z)}. A Bayesian technique is also employed to directly determine
the correlation lengths, marginalizing over the slope, $\delta$,
subject to the weak prior that it is in the range $0.2 \le \delta \le
1.8$. This method is desirable since, despite the large number of
clusters, it is difficult to simultaneously constrain both amplitude
and slope. Figure \ref{Fig: ACF} shows both the frequentist fits and
the Bayesian likelihood functions in $r_0$.

In this highly clustered population, sample (or cosmic) variance
dominates over simple Poisson errors. We therefore calculate the error
bars using 100 bootstrap resamplings with replacement. To test for
possible systematics across the field we divided the field into
halves, once north--south and once east--west, and for each subfield
we computed the ACF. The results for all half--fields agree within
1\,$\sigma$, indicating that we are not adversely affected by an
unidentified bias in the spatial selection function.

Bootstrap simulations of the cluster detection process indicate that
the spurious fraction is less than 10\% at all redshifts (E07), and
spectroscopic observations indicate it is likely much lower.
Conservatively assuming 5--10\% of the sample is indeed spurious, and
that these are uncorrelated, then at most we are underestimating the
clustering by $\approx$ 11--23\%.

In this work we adopt the $r_0$ values from the Bayesian fits, though
the fits from both methods, presented in Table \ref{Tab: Fits}, are
completely consistent.  The correlation amplitude in our low redshift
bin at $z \approx 0.5$ is in excellent agreement with the only other
measurement at this redshift \citep{gonzalez02}.  Our high redshift
measurement, at $z \approx 1$, is the first to probe structure on the
largest scales in the first half of Universe.

\begin{figure}[bthp]
\epsscale{1.2}
\plotone{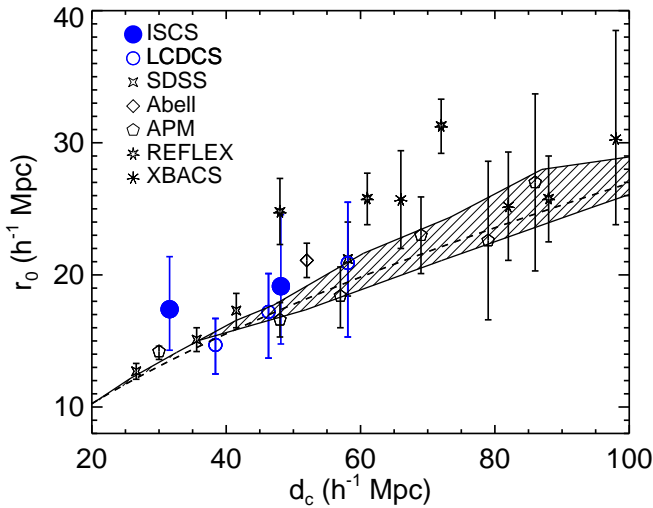}  
\caption{Plot of $r_0$ vs.\ $d_c$ for the present sample, along with
  several measurements taken from the compilation of
  \citet{bahcall03}.  Our results (filled blue circles) are consistent
  with LCDM predictions between $0 <z< 1.5$ (hashed region), with the
  parametrization of \citet[dashed line]{younger05}, as well as with
  the most distant measurement prior to the present one, the LCDCS at
  $z \sim 0.5$ (open blue circles).  Data are from
  \cite{gonzalez02,abadi98,croft97,peacock92,bahcall83,collins00,lee&park99}.
}
\label{Fig: d_c}
\end{figure}

The space densities for these samples and the mean intercluster
distances, $d_c$, are also presented in Table \ref{Tab: Fits}.  The
relationship between clustering amplitude and $d_c$ predicted in a
concordance cosmology, and observed in practice \citep[and references
therein]{bahcall03}, is only weakly dependent on redshift.  As shown
in Figure \ref{Fig: d_c}, the ISCS samples are quite consistent with
the LCDM predictions between $0 < z < 1.5$ (hashed region) from
\citet{younger05}.

\begin{deluxetable*}{ccccccccc}
\tablecolumns{8}
\tabletypesize{\normalsize}
\tablecaption{Amplitude of Correlations for ISCS Clusters\label{Tab: Fits}}
\tablewidth{0pt}
\tablehead{
\colhead{$\Delta z$} & \colhead{$z_{\mbox{\small eff}}$} &
\colhead{$N$} & \colhead{$\log A_\omega$\tablenotemark{a}} & \colhead{$r_0$\tablenotemark{b}}& 
\colhead{$\delta$\tablenotemark{b}}&
\colhead{$r_0$\tablenotemark{c}} & \colhead{$\bar{n}$}&\colhead{$d_c$} \\ 
&&&&\colhead{(\hmpc)}&&\colhead{(\hmpc)}&\colhead{($10^{-6}$ $h^{3}$ Mpc$^{-3}$)}&\colhead{(\hmpc)}}
\startdata
0.25 -- 0.75 & 0.53 & 136 & $-1.37 \pm ^{0.18}_{0.31}$& $ 18.70 \pm ^{4.31}_{5.66}  $& $ 1.00 \pm ^{0.29}_{0.46}   $ &$ 17.40\pm ^{3.98}_{3.10}  $ & $31.8 \pm 2.7$ &  $31.6 \pm 0.9$\\ \\
0.75 -- 1.50 & 0.97 & 141 & $-1.73 \pm ^{0.52}_{0.23}$& $ 18.64 \pm ^{5.18}_{7.66}   $& $ 1.21 \pm ^{0.29}_{0.94}  $ &$ 19.14\pm ^{5.65}_{4.56}  $ &  $8.96 \pm 0.76$ &  $48.1 \pm 1.4 $\\
\enddata 
\tablenotetext{a}{Formal fits for $r_0$ and $\delta$ were computed
  directly from the data.  The error in $A_\omega$ corresponds to the
  error in $r_0$ at the best--fit $\delta$.}
\tablenotetext{b}{Best--fit parameters from frequentist fits.}
\tablenotetext{c}{Best--fit $r_0$ from Bayesian marginalization.}
\end{deluxetable*}

\section{Discussion}
\label{Sec: discussion}


\begin{figure}[htbp]
\epsscale{1.1}
\plotone{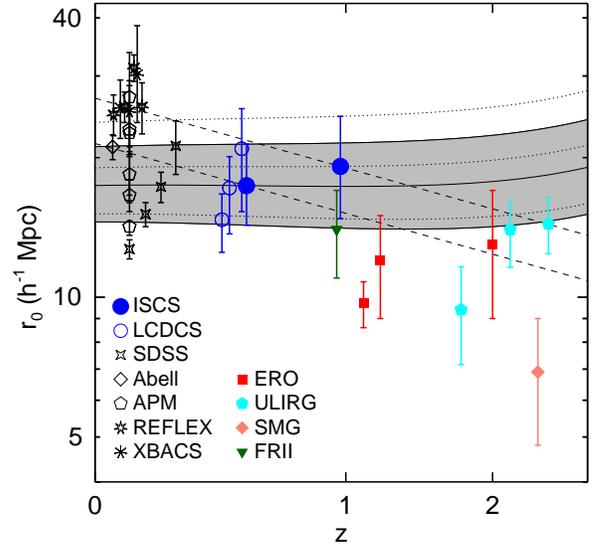}  
\caption{Comoving correlation lengths for the ISCS clusters, along
  with other cluster samples at lower redshift (offset slightly in z
  for clarity) and highly clustered galaxy samples at higher redshifts
  (see the text for a full description and references).  Two
  evolutionary models are overplotted, including the \citet{fry96}
  biased structure formation model (shaded region and dotted lines for
  $z_{\mbox{\scriptsize eff}}=0.53$ and $0.97$, respectively), and a
  simple stable clustering model (dashed lines). }
\label{Fig: theory and evolution}
\end{figure}

A key theoretically predictable cluster observable is the correlation
function as a function of halo mass.  In simulations the halo mass,
\mtwo, is defined as the mass inside the radius at which the mean
overdensity is 200 times the critical density.  We compare our
clustering results with the \citet{younger05} analysis of the
\citet{hopkins05} high--resolution cosmological simulation, which had
a 1500 \hmpc\ box length, an individual particle mass of $1.8 \times
10^{11} \msun$, and a power spectrum normalization of $\sigma_8 =
0.84$.  We infer that the ISCS cluster sample has average $\log
[M_{\mbox{\scriptsize 200}}/M_\odot]$ masses of $\sim
13.9^{+0.3}_{-0.2}$ and $\sim 13.8^{+0.2}_{-0.3}$ at
$z_{\mbox{\scriptsize eff}}=0.53$ and $0.97$, respectively.  Direct
dynamical masses for the 10 $z>1$ clusters presented in E07 yield a
largely consistent distribution of masses (Gonzalez \etal\ in prep).


The observed constancy of $r_0$ for massive galaxy clusters out to
$z=1$ is a robust confirmation of a key prediction from numerical
simulations, reflecting the relative constancy of the mass hierarchy
of clusters with redshift \citep{younger05}.  That is, the N most
massive clusters at one epoch roughly correspond to the N most massive
clusters at a later epoch, and therefore have similar clustering.

In Figure \ref{Fig: theory and evolution} we plot a compilation of
recent clustering amplitudes for various cluster surveys, as well as
for highly clustered galaxy populations, including FRII radio galaxies
\citep{overzier03}, EROs \citep{brown05,daddi04,daddi01}, ULIRGs
\citep{farrah06, magliocchetti07}, and SMGs \citep{blain04}.  Since
optically--selected QSOs are more modestly clustered
\citep{porciani04, croom05, myers06, coil07}, and therefore reside in
considerably less massive halos ($\sim 10^{12}-10^{13} M_\odot$) than
the ISCS clusters, they are not included in Figure \ref{Fig: theory
  and evolution}.

Following \citet{moustakas&somerville02}, we overplot the halo
conserving model of \citet{fry96} normalized to our two measurements
in order to explore possible evolutionary connections with structures
at other redshifts.  The shaded area (dotted lines) shows the
1\,$\sigma$ region for the $z_{\mbox{\scriptsize eff}}=0.53$ ($0.97$)
measurement.  In this model, representative of a class of merger-free
biased structure formation models, the ISCS clusters will evolve into
typical present--day massive clusters, such as those in the SDSS, APM
or Abell surveys.  In the stable--clustering picture, in which
clustering is fixed in physical coordinates \citep[dashed
lines]{groth&peebles77}, the $z_{\mbox{\scriptsize eff}}=0.97$ ISCS
clusters grow into the most massive clusters in the local Universe,
typically identified in X--ray surveys.

Most of the plotted high redshift galaxy clustering measurements are
rather uncertain due to both small number statistics and poorly known
redshift distributions.  Mindful of this caveat, we observe that FRII
radio galaxies, some ERO samples, and $z\sim2$ ULIRGs have clustering
consistent with that seen in the ISCS clusters in either model.
Clearly these populations trace very massive halos ($\ga 10^{13}
M_\odot$).  As a caution against overinterpretation, however, we note
that only ULIRGs have space densities similar to the present cluster
samples, a prerequisite for drawing evolutionary connections from
these particular models.  Thus the present work offers a measure of
support for recent studies \citep[\eg][]{magliocchetti07} indicating
that ULIRGs may be associated with, or progenitors of, groups or
low--mass clusters.

\section{Conclusions}
\label{Sec: conclusion}

By deprojecting the angular correlation function measured in redshift
bins spanning $z=0.25$ to $z=1.5$, we have determined the real--space
clustering amplitudes for ISCS clusters at $\left<z\right>=0.53$ and
$\left<z\right>=0.97$ to be $r_0 = 17.40^{+3.98}_{-3.10}$ and
$r_0=19.14^{+5.65}_{-4.56}$ \hmpc, respectively.  These measurements
are consistent with the relation between correlation amplitude and
mean intercluster distance predicted by LCDM.  The ISCS clusters have
total masses exceeding $10^{14} M_\odot$ and will evolve into very
massive clusters by the present day.

\acknowledgements  

Based in part on observations made with the {\it Spitzer Space
  Telescope}, operated by the Jet Propulsion Laboratory, California
Institute of Technology, under a contract with NASA. This paper made
use of data from the NDWFS, which was supported by NOAO, AURA, Inc.,
and the NSF. We thank A.~Dey, B.~Jannuzi and the entire NDWFS survey
team.  We also thank J.~Younger and P.~Hopkins for providing their
simulation results, C.\ Porciani and K.\ Blindert for useful
discussions, and the anonymous referee for a very helpful report.
SAS's work was performed under the auspices of the U.S. DoE, National
Nuclear Safety Administration by the University of California, LLNL
under contract No.~W-7405-Eng-48.

\bibliographystyle{astron3}
\bibliography{bibfile}

\end{document}